\documentclass[12pt,a4paper]{article}
\usepackage{amssymb}
\usepackage{amsfonts}

\newtheorem{theorem}{Theorem}[section]
\newtheorem{proposition}[theorem]{Proposition}

\newcommand{\ba}{\begin{array}}
\newcommand{\ea}{\end{array}}
\newcommand{\pa}{\partial}

\newcommand{\La}{\Lambda}
\newcommand{\ep}{\epsilon}
\newcommand{\no}{\nonumber}
\newcommand{\mL}{\mathcal{L}}
\newcommand{\mM}{\mathcal{M}}

\newcommand{\be}{\begin{equation}}
\newcommand{\ee}{\end{equation}}
\newcommand{\bea}{\begin{eqnarray}}
\newcommand{\eea}{\end{eqnarray}}
\newcommand{\beaa}{\begin{eqnarray*}}
\newcommand{\eeaa}{\end{eqnarray*}}

\newcommand{\nn}{\nonumber}

\begin{document}

\title{Generalized dKP: Manakov-Santini hierarchy and its waterbag reduction}
%\author{}
\author{
L.V. Bogdanov\thanks
{L.D. Landau ITP, Kosygin str. 2,
Moscow 119334, Russia
%, E-mail
%leonid@landau.ac.ru
}, Jen-Hsu Chang$^\ddag$ \thanks {Department of Computer Science,
 National Defense University,
 Taoyuan, Taiwan,
E-mail: jhchang@ndu.edu.tw, $\ddag$ corresponding author. }~ and
Yu-Tung Chen$^\dag$ }

\maketitle

\begin{abstract}
We study Manakov-Santini equation, starting from Lax-Sato form of associated hierarchy. The
waterbag reduction for Manakov-Santini hierarchy is introduced.
Equations of reduced hierarchy are derived.
We construct new coordinates transforming non-hydrodynamic
evolution of waterbag reduction to non-homogeneous Riemann
invariants form of hydrodynamic type.
\end{abstract}
 \maketitle
%%%%%%%%%%%%%%%%%%%%%%%%%%%%%%%%%%%%%%%%%%%%%%%%%%%%%%%%%
%
\section{Introduction}
%
%%%%%%%%%%%%%%%%%%%%%%%%%%%%%%%%%%%%%%%%%%%%%%%%%%%%%%%%%
In this paper we study an integrable system introduced recently by
Manakov and Santini \cite{MS06} (see also \cite{MS07,MS08}). This
system is connected with commutation of general 2-dimensional
vector fields (containing derivative on spectral variable).
Reduction to Hamiltonian vector fields leads to the well-known
dispersionless KP (or Khokhlov-Zabolotskaya) equation.
Alternatively, a natural reduction to 1-dimensional vector fields
reduces Manakov-Santini system  to the equation introduced by
Pavlov \cite{Pavlov03} (see also \cite{Dun04,MS02,MS04}). Using
general construction of the works \cite{BDM06,BDM07}, we introduce
the hierarchy for Manakov-Santini system in Lax-Sato form and
generating equation for it (the hierarchy in terms of recursion
operator was introduced in \cite{MS07}). We introduce waterbag
ansatz for Manakov-Santini hierarchy and derive equations of the
reduced hierarchy. Using rational form of the G function (see
below), one can introduce new coordinates such that the
non-hydrodynamic evolution of waterbag reduction transforms to
non-homogeneous Riemann invariants form of hydro-dynamic type. \\
\indent This paper is organized as follows.
%%%%%%%%%%%%%%%%%%%%%%%%%%%
% Section 2
%%%%%%%%%%%%%%%%%%%%%%%%%%%
In section 2, GdKP hierarchy is described, connection to Manakov-Santini system is
demonstrated.
%%%%%%%%%%%%%%%%%%%%%%%%%%%
% Section 3
%%%%%%%%%%%%%%%%%%%%%%%%%%%
In section 3, waterbag reduction for Manakov-Santini hierarchy is introduced,
equations of reduced hierarchy are derived (in non-hydrodynamic form).
%%%%%%%%%%%%%%%%%%%%%%%%%%%
% Section 4
%%%%%%%%%%%%%%%%%%%%%%%%%%%
In section 4, we introduce new coordinates transforming the evolution
of waterbag reduction to non-homogeneous Riemann
invariants form of hydro-dynamic type. The examples are given.
%%%%%%%%%%%%%%%%%%%%%%%%%%%
% Section 5
%%%%%%%%%%%%%%%%%%%%%%%%%%%
Section 5 is devoted to the concluding remarks.

%%%%%%%%%%%%%%%%%%%%%%%%%%%%%%%%%%%%%%%%%%%%%%%%%%%%%%%%%
%
\section{Generalized dKP hierarchy}
%
%%%%%%%%%%%%%%%%%%%%%%%%%%%%%%%%%%%%%%%%%%%%%%%%%%%%%%%%%
To introduce generalized dKP (Manakov-Santini) hierarchy, we use general construction of the works
\cite{BDM06,BDM07}.
The hierarchy is described by the Lax-Sato equations
\begin{eqnarray}
\frac{\pa\psi}{\pa t_n}
= A_n\frac{\pa\psi}{\pa x} - B_n\frac{\pa\psi}{\pa p},
\qquad \psi=\left(\ba{c} \mL\\ \mM \ea\right),
\label{G-dKP}
\end{eqnarray}
or, equivalently, by the generating equation
\be
(J_0^{-1}\mathrm{d}\mL\wedge \mathrm{d}\mM)_-=0,
\ee
where $A_n\equiv(J_0^{-1}\pa\mL^n/\pa p)_+$, $B_n\equiv(J_0^{-1}\pa\mL^n/\pa x)_+$
with the Lax and Orlov operators $\mL(p), \mM(p)$ being the Laurent series
\begin{eqnarray}
\mL &=& p+\sum_{n=1}^{\infty}u_n(x)p^{-n},
\label{L}\\
\mM &=& \sum_{n=1}^{\infty}nt_n\mL^{n-1}+\sum_{n=1}^{\infty}v_n(x)\mL^{-n}.
\label{M}
\end{eqnarray}
Here $(\cdots)_+$ ($(\cdots)_-$)
denote respectively the projection on the polynomial part (negative powers), and $J_0$ is defined by
\begin{eqnarray*}
J_0 &=& \frac{\pa \mL}{\pa p}\frac{\pa \mM}{\pa x}-\frac{\pa \mL}{\pa x}\frac{\pa \mM}{\pa p} \\
    &=& \frac{\pa \mL}{\pa p}\left(\left.\frac{\pa \mM}{\pa \mL}\right|_{t_n,v_n\ \rm{fixed}}
        \frac{\pa \mL}{\pa x}
         +\left.\frac{\pa \mM}{\pa x}\right|_{\mL\ \rm{fixed}}\right)
         -\frac{\pa \mL}{\pa x}\left.\frac{\pa \mM}{\pa \mL}\right|_{t_n,v_n\ \rm{fixed}}
        \frac{\pa \mL}{\pa p} \\
    &=& \frac{\pa \mL}{\pa p}\left.\frac{\pa \mM}{\pa x}\right|_{\mL\ \rm{fixed}}
     = 1+v_{1x}p^{-1}+(v_{2x}-u_1)p^{-2}+\cdots.
\end{eqnarray*}
We list some of $A_n$ and $B_n$ as follows
\begin{eqnarray}
A_1 &=& 1, \no\\
A_2 &=& 2p-2v_{1x}, \no\\
A_3 &=& 3p^2-3v_{1x}p+6u_1+3(v_{1x})^2-3v_{2x}, \no\\
A_4 &=& 4p^3-4v_{1x}p^2+(12u_1+4(v_{1x})^2-4v_{2x})p \no\\
     && +12u_2-4v_{3x}+8v_{1x}v_{2x}-4(v_{1x})^3-8u_1v_{1x},
\label{A}
\end{eqnarray}
and
\begin{eqnarray}
B_1 &=& 0, \no\\
B_2 &=& 2u_{1x}, \no\\
B_3 &=& 3u_{1x}p - 3u_{1x}v_{1x}+3u_{2x}, \no\\
B_4 &=& 4u_{1x}p^2+(4u_{2x}-4u_{1x}v_{1x})p \no\\
     && +4u_{1x}(4u_1+(v_{1x})^2-v_{2x})-4u_{2x}v_{1x}+4u_{3x}.
\label{B}
\end{eqnarray}
The $t_1$ flow of the generalized dKP hierarchy (\ref{G-dKP}) says that the
dependence on $t_1$ and $x$ appear in the linear combination $t_1+x$.
%%%%%%%%%%%%%%%%%%%%%%%%%%%%%%%%%%%
% Proposition 1
%%%%%%%%%%%%%%%%%%%%%%%%%%%%%%%%%%%
\begin{proposition}
The compatibility of the commuting flow $[\pa_{t_m}, \pa_{t_n}]\psi=0$
requires $A_n,B_n$ to satisfy
\begin{eqnarray}
\pa_{t_m}A_n-\pa_{t_n}A_m &=& \langle A_m, A_n\rangle_x + B_nA_{mp} - B_mA_{np}, \no\\
\pa_{t_m}B_n-\pa_{t_n}B_m &=& \langle B_n, B_m\rangle_p + A_mB_{nx} - A_nB_{mx},
\label{An-Bn}
\end{eqnarray}
where $\langle U,V \rangle_i:=U(\pa_i V)-(\pa_i U)V$.
\end{proposition}
\textit{Proof.}
Substituting (\ref{G-dKP}) into $\pa_{t_m}\pa_{t_n}\psi=\pa_{t_n}\pa_{t_m}\psi$,
and comparing the coefficients of independent variables $\psi_x$ and $\psi_p$ respectively
to the both sides, we obtain (\ref{An-Bn}). $\square$
%%%%%%%%%%%%%%%%%%%%%%%%%%%%%%%%%%%

The evolution of $\mL,\mM$ with respect to $t_2=y$ in (\ref{G-dKP}) are given by
\begin{eqnarray}
\frac{1}{2}\frac{\pa \mL}{\pa y} &=& (p-v_{1x})\frac{\pa \mL}{\pa x} - u_{1x}\frac{\pa \mL}{\pa p},
\label{L_t2} \\
\frac{1}{2}\frac{\pa \mM}{\pa y} &=& (p-v_{1x})\frac{\pa \mM}{\pa x} - u_{1x}\frac{\pa \mM}{\pa p}.
\label{M_t2}
\end{eqnarray}
Using the convention $(\sum_na_np^n)_{[s]}=a_s$ for a formal Laurent series,
then from Eq.(\ref{L_t2}) we have
\begin{eqnarray}
\frac{1}{2}u_{1y}
&=& \Big((p-v_{1x})\mL_x-u_{1x}\mL_p\Big)_{[-1]}
 = u_{2x} - v_{1x}u_{1x},
\label{u1_y}\\
\frac{1}{2}u_{2y}
&=& \Big((p-v_{1x})\mL_x-u_{1x}\mL_p\Big)_{[-2]}
= u_{3x} - v_{1x}u_{2x}+u_1u_{1x}
\label{u2_y},
\end{eqnarray}
On the other hand, the expression of Eq.(\ref{M_t2}) together with (\ref{L_t2}) gives
\[
 \mL + \frac{1}{2}\sum_{n=1}^{\infty}v_{ny}\mL^{-n}
= (p-v_{1x})\left(1+\sum_{n=1}^{\infty}v_{nx}\mL^{-n}\right).
\]
Comparing the coefficients of powers $p^{-1}$ and $p^{-2}$ to the above, we have
\begin{eqnarray}
v_{2x} &=& u_1 + v_{1x}^2 +\frac{1}{2}v_{1y},
\label{v2_x} \\
v_{3x} &=& u_2 + \frac{1}{2}v_{2y}+ u_1v_{1x} +v_{1x}v_{2x}.
\label{v3_x}
\end{eqnarray}
Similarly, the evolution of $\mL,\mM$ w.r.t. $t_3=t$ are given by
\begin{eqnarray}
\frac{1}{3}\frac{\pa \mL}{\pa t} &=& \left(p^2-v_{1x}p+u_1-\frac{1}{2}v_{1y}\right)\frac{\pa \mL}{\pa x}
                                   -\left(u_{1x}p+\frac{1}{2}u_{1y}\right)\frac{\pa \mL}{\pa p},
\label{L_t3} \\
\frac{1}{3}\frac{\pa \mM}{\pa t} &=& \left(p^2-v_{1x}p+u_1-\frac{1}{2}v_{1y}\right)\frac{\pa \mM}{\pa x}
                                   -\left(u_{1x}p+\frac{1}{2}u_{1y}\right)\frac{\pa \mM}{\pa p},
\label{M_t3}
\end{eqnarray}
Then the $t$-flow of $u_1$ can be read by Eq.(\ref{L_t3}) by taking the coefficient of $p^{-1}$:
\begin{eqnarray}
\frac{1}{3}u_{1t}
&=& u_{3x}-v_{1x}u_{2x}+(u_1-\frac{1}{2}v_{1y})u_{1x}+u_1u_{1x}, \no \\
&=& \frac{1}{2}u_{2y}-\frac{1}{2}u_{1x}v_{1y}+u_1u_{1x},
\label{u1_t}
\end{eqnarray}
where we have used (\ref{u2_y}) to reach the second line.
Also, the expression of Eq.(\ref{M_t3}) together with (\ref{L_t3}) gives
\[
 \mL^2+\frac{1}{3}\sum_{n=1}^{\infty}v_{nt}\mL^{-n}
=\left(p^2-v_{1x}p+u_1-\frac{1}{2}v_{1y}\right)\left(1+\sum_{n=1}^{\infty}v_{nx}\mL^{-n}\right),
\]
in which the coefficient of $p^{-1}$ gives
\begin{equation}
\frac{1}{3}v_{1t} = -u_2 + \frac{1}{2}v_{2y}+u_1v_{1x}-\frac{1}{2}v_{1x}v_{1y},
\label{v1_t}
\end{equation}
where we have used Eq.(\ref{v3_x}).
Now differentiating Eqs.(\ref{u1_t}), (\ref{v1_t}) respectively with respect to $x$
and eliminating $u_{2x}$ and $v_{2x}$ by Eqs.(\ref{u1_y}) and (\ref{v2_x}),
we obtain the following two coupled equations
for $u_1:=u$ and $v_1=v$:
\begin{eqnarray}
\frac{1}{3}u_{xt} &=& \frac{1}{4}u_{yy}+(uu_x)_x+\frac{1}{2}v_xu_{xy}-\frac{1}{2}u_{xx}v_y,
\nn\\
\frac{1}{3}v_{xt} &=& \frac{1}{4}v_{yy}+uv_{xx}+\frac{1}{2}v_xv_{xy}-\frac{1}{2}v_{xx}v_y.
\label{MSeq}
\end{eqnarray}
Eq.(\ref{MSeq}) is the so called \emph{Manakov-Santini equation} \cite{MS06,MS07,MS08}.
The Lax pair for this equation is defined by linear equations (\ref{L_t2},\ref{M_t2}) and
(\ref{L_t3},\ref{M_t3}).
Notice that for $v=0$ reduction, the system reduces to the dKP equation
\begin{equation}
 \frac{1}{3}u_{xt} = \frac{1}{4}u_{yy}+(uu_x)_x.
\label{dKP-eq}
\end{equation}
Respectively, $u=0$ reduction gives an equation \cite{Pavlov03} (see also \cite{Dun04,MS02,MS04})
\begin{equation}
 \frac{1}{3}v_{xt} = \frac{1}{4}v_{yy}+\frac{1}{2}v_xv_{xy} - \frac{1}{2}v_{xx}v_y.
\label{Pavlov}
\end{equation}
%%%%%%%%%%%%%%%%%%%%%%%%%%%%%%%%%%%
% Proposition 2
%%%%%%%%%%%%%%%%%%%%%%%%%%%%%%%%%%%
\begin{proposition}
Equation (\ref{G-dKP}) can be written in Hamilton-Jacobi type equation
\begin{equation}
 \left.\frac{\pa p(\mL)}{\pa t_n}\right|_{\mL\ \rm{fixed}}
= \left.A_n(p(\mL))\frac{\pa p(\mL)}{\pa x}\right|_{\mL\ \rm{fixed}} + B_n(p(\mL)),
\end{equation}
where $A_n(p)=(J_0^{-1}\pa \mL^n/\pa p)_+$ and $B_n(p)=(J_0^{-1}\pa \mL^n/\pa x)_+$.
\end{proposition}
\textit{Proof.}
By taking into account the partial derivatives with respect to $t_n$ for fixed $p$ or $\mL$,
it is easy to show that
\[
 \frac{\pa p}{\pa t_n}=0
=\left.\frac{\pa p(\mL)}{\pa t_n}\right|_\mL+\frac{\pa p(\mL)}{\pa \mL}\frac{\pa \mL}{\pa t_n},
\]
or
\begin{equation}
 \left.\frac{\pa p(\mL)}{\pa t_n}\right|_\mL = -\frac{\pa p(\mL)}{\pa \mL}\frac{\pa \mL}{\pa t_n}.
\label{pLeq}
\end{equation}
Using (\ref{G-dKP}), and (\ref{pLeq}) with $n=1$, we have
\[
\left.\frac{\pa p(\mL)}{\pa t_n}\right|_{\mL\ \rm{fixed}}
=-\frac{\pa p(\mL)}{\pa \mL}\left(A_n(p)\frac{\pa \mL}{\pa x}-B_n(p)\frac{\pa \mL}{\pa p}\right)
=\left.A_n(p(\mL))\frac{\pa p(\mL)}{\pa x}\right|_{\mL\ \rm{fixed}} + B_n(p(\mL)).\,\square
\]
%%%%%%%%%%%%%%%%%%%%%%%%%%%%%%%%%%%
%
%%%%%%%%%%%%%%%%%%%%%%%%%%%%%%%%%%%
% Proposition 3
%%%%%%%%%%%%%%%%%%%%%%%%%%%%%%%%%%%
\begin{proposition}\label{J-G}
The function $J_0=\pa_p\mL\pa_x\mM-\pa_x\mL\pa_p\mM$ and its inverse $G=J_0^{-1}$ satisfy
\begin{eqnarray}
 \pa_{t_n}J_0 &=& \left(A_nJ_0\right)_x-\left(B_nJ_0\right)_p,
\label{J0eq}\\
 \pa_{t_n}G &=& \langle A_n,G \rangle_x - \langle B_n,G\rangle_p,
\label{Geq}
\end{eqnarray}
where $\langle U,V \rangle_i:=U(\pa_i V)-(\pa_i U)V$.
\end{proposition}
\textit{Proof.}
Using the $t_n$-flows of $\mL,\mM$ in (\ref{G-dKP}) and the definition of $J_0$, we have
\begin{eqnarray*}
 \pa_{t_n}J_0
&=& (\mL_p)_{t_n}\mM_x+\mL_p(\mM_x)_{t_n}- (\mL_x)_{t_n}\mM_p-\mL_x(\mM_p)_{t_n}, \\
&=& -B_{np}J_0 + A_{nx}J_0 +A_nJ_{0x} -B_nJ_{0p}, \\
&=& (A_nJ_0)_x - (B_nJ_0)_p.
\end{eqnarray*}
Moreover, substituting $J_0=G^{-1}$ into the above we obtain (\ref{Geq}). $\square$ \\
%This immediately implies that Eqs.(\ref{J0eq}) and (\ref{Geq}) are
%equivalent to the generalized dKP hierarchy.
As we will see, Proposition \ref{J-G} can provide a crucial way to
determine the hierarchy flows.
%%%%%%%%%%%%%%%%%%%%%%%%%%%%%%%%%%%%%%%%%%%%%%%%%%%%%%%%%
%
\section{Waterbag-type reduction}
%
%%%%%%%%%%%%%%%%%%%%%%%%%%%%%%%%%%%%%%%%%%%%%%%%%%%%%%%%%
Consider the waterbag-type reduction of the generalized dKP
hierarchy represented by \cite{BDM06}
\begin{eqnarray}
\mL &=& p+\sum_{i=1}^{N}\ep_i\log(p-U_i),
\label{waterbagL}\\
\mM &=& \sum_{n=1}^{\infty}nt_n\mL^{n-1}+\sum_{i=1}^{M}\delta_i\log(p-V_i),
\label{waterbagM}
\end{eqnarray}
where $\ep_i$ and $\delta_i$ are assumed to satisfy \be
\sum_{i=1}^{N}\ep_i=\sum_{i=1}^{M}\delta_i=0. \label{zero} \ee The
ansatz (\ref{waterbagL},\ref{waterbagM}) is consistent with the
dynamics defined by Manakov-Santini hierarchy (\ref{G-dKP}), i.e.,
the form of ansatz is preserved by the dynamics. Condition
(\ref{zero}) guarantees that expansion of $\mL$, $\mM$ at infinity
is of the form (\ref{L},\ref{M}). Reduced hierarchy is represented
as infinite set of (1+1)-dimensional systems of equations for the
functions $U_i$, $V_i$, which are obtained by the substitution of
ansatz (\ref{waterbagL},\ref{waterbagM}) to equations of
Manakov-Santini hierarchy (\ref{G-dKP}).

Let us consider first flows of reduced hierarchy. For expansion of $\mL$, $\mM$ at infinity
from (\ref{waterbagL},\ref{waterbagM}) we get
\bea
&&
\mL=p-\sum_{n=1}^{\infty} \left(\sum_{i=1}^{N}\ep_i\frac{U_i^n}{n}\right) p^{-n},
\label{wL}
\\&&
\mM =\sum_{n=1}^{\infty}nt_n\mL^{n-1}-\sum_{n=1}^{\infty} \left( \sum_{i=1}^{M}\delta_i\frac{V_i^n}{n} \right)  p^{-n}.
\label{wM}
%u_n=- \sum_{i=1}^{N}\ep_i\frac{U_i^n}{n},\quad
%v_n=- \sum_{i=1}^{N}\delta_i\frac{V_i^n}{n}.
\eea
Comparing these expansions with formulae (\ref{L},\ref{M}), we come to the conclusion that
$u_n=- \sum_{i=1}^{N}\ep_i\frac{U_i^n}{n}$. To calculate $v_n$, we should invert the series (\ref{wL}) to find $p(\mL)$ that can be done recursively, and substitute $p(\mL)$ to  (\ref{wM}) . For the first coefficients $u_n$,  $v_n$ we get
\beaa
&&
u_1=-\sum_{i=1}^{N}\ep_i U_i, \quad u_2=- \frac{1}{2} \sum_{i=1}^{N}\ep_i U_i^2,
\\
&&
v_1=- \sum_{i=1}^{M}\delta_i V_i, \quad v_2=-\frac{1}{2}\sum_{i=1}^{M}\delta_i V_i^2.
\eeaa
Substituting these expressions to relations (\ref{A}), (\ref{B}) and using equations (\ref{G-dKP}),
we obtain equations of reduced hierarchy.
Equations of the flow corresponding to $y=t_2$ read
\bea
&&
\partial_y U_k=
\Big(2U_k + \partial_x\sum_{i=1}^{M}\delta_i V_i\Big)\partial_x U_k - 2\partial_x\Big(\sum_{i=1}^{N}\ep_i U_i\Big),
\nn\\
&&
\partial_y V_k=
\Big(2V_k+\partial_x\sum_{i=1}^{M}\delta_i V_i\Big)\partial_x V_k - 2\partial_x\Big(\sum_{i=1}^{N}\ep_i U_i\Big).
\label{y-flow}
\eea
For the flow corresponding to $t=t_3$ we get
\bea
&&
\partial_t U_k=
\left(3 U_k^2 + 3 U_k\partial_x\sum_{i=1}^{M}\delta_i V_i
-6\sum_{i=1}^{N}\ep_i U_i + 3\bigl(\partial_x\sum_{i=1}^{M}\delta_i V_i\bigr)^2
+3 \partial_x\sum_{i=1}^{M}\delta_i\frac{V_i^2}{2}\right)\partial_x U_k
-
\nn\\
&&\qquad
\left(3 U_k \partial_x\sum_{i=1}^{N}\ep_i U_i+ 3(\partial_x\sum_{i=1}^{N}\ep_i U_i)
(\partial_x\sum_{i=1}^{M}\delta_i V_i) + 3\partial_x\sum_{i=1}^{N}\ep_i\frac{U_i^2}{2}\right),
\nn\\
&&
\partial_t V_k=
\left(3 V_k^2 + 3 V_k\partial_x\sum_{i=1}^{M}\delta_i V_i
-6\sum_{i=1}^{N}\ep_i U_i + 3\bigl(\partial_x\sum_{i=1}^{M}\delta_i V_i\bigr)^2
+3 \partial_x\sum_{i=1}^{M}\delta_i\frac{V_i^2}{2}\right)\partial_x V_k
-
\nn\\
&&\qquad
\left(3 V_k \partial_x\sum_{i=1}^{N}\ep_i U_i+ 3(\partial_x\sum_{i=1}^{N}\ep_i U_i)
(\partial_x\sum_{i=1}^{M}\delta_i V_i) + 3\partial_x\sum_{i=1}^{N}\ep_i\frac{U_i^2}{2}\right).
\label{t-flow}
\eea
A common solution to the systems  (\ref{y-flow}), (\ref{t-flow}) gives a solution
to Manakov-Santini equation (\ref{MSeq}) defined as
$u=-\sum_{i=1}^{N}\ep_i U_i$, $v=-\sum_{i=1}^{M}\delta_i V_i$.

%%%%%%%%%%%%%%%%%%%%%%%%%%%%%%%%%%%%%%%%%%%%%%%%%%%%%%%%%%%%%%%%%
\section{Diagonal form of reduced hierarchy}
%%%%%%%%%%%%%%%%%%%%%%%%%%%%%%%%%%%%%%%%%%%%%%%%%%%%%%%%%%%%%%%%%
For the waterbag reduction (\ref{waterbagL}, \ref{waterbagM})
one can show that the $G$ function can be expressed in the following form
\begin{equation}
 G=J_0^{-1}=\frac{\prod_{i=1}^{N}(p-U_i)\prod_{j=1}^{M}(p-V_j)}
                 {F(U_n,U_{nx},V_m,V_{mx};p)},
 \qquad n=1,\ldots,N;\,m=1,\ldots,M,
\label{G-Wb}
\end{equation}
where the function $F$ in denominator is a polynomial of $p$ with degree $N+M$.
In general, $F$ can also be factorized into $\prod_{k=1}^{N+M}(p-W_k)$,
for which $W_k=W_k(U_n,U_{nx},V_m,V_{mx})$ are roots of $J_0$.
We like to mention here that the derivatives $U_{nx}, V_{mx}$ can be
inversely expressed as function of the form
\begin{equation}
 U_{nx}=f_n(U_i,V_j,W_k), \quad V_{mx}=g_m(U_i,V_j,W_k).
\label{UnVm}
\end{equation}
Therefore, we have
\begin{equation}
 J_0 = \frac{\prod_{k=1}^{N+M}(p-W_k)}{\prod_{i=1}^{N}(p-U_i)\prod_{j=1}^{M}(p-V_j)},
 \qquad n=1,\ldots,N;\,m=1,\ldots,M.
\label{J0-Wb}
\end{equation}
As the result, the evaluation of $G$ at $U_i$ or $V_i$, i.e., $G(p=U_i)=0$ or $G(p=V_i)=0$
shows that Eq.(\ref{Geq}) can be written into the following evolution equations of $U_i$, $V_i$:
\begin{eqnarray}
 \frac{\pa U_i}{\pa t_n} &=& A_n(p=U_i)\frac{\pa U_i}{\pa x} + B_n(p=U_i),
\label{U-tn} \\
 \frac{\pa V_i}{\pa t_n} &=& A_n(p=V_i)\frac{\pa V_i}{\pa x} + B_n(p=V_i).
\label{V-tn}
\end{eqnarray}
Similarly, Eq.(\ref{J0eq}) with $J_0(p=W_i)=0$ gives rise
\begin{equation}
 \frac{\pa W_i}{\pa t_n} = A_n(p=W_i)\frac{\pa W_i}{\pa x} + B_n(p=W_i).
\label{W-tn}
\end{equation}
In summary, combining (\ref{U-tn}), (\ref{V-tn}), (\ref{W-tn}) and
replacing those $U_{nx}$'s and $V_{mx}$'s in $A_n, B_n$ with the
transformations (\ref{UnVm}),
we obtain the \emph{non-homogeneous Riemann invariant form} as
\begin{equation}
\pa_{t_n}R_i = A_n(p=R_i)R_{ix} + B_n(p=R_i), \quad i=1,\ldots,2N+2M,
\label{N-Rie-inv}
\end{equation}
for which $(R_1,\ldots,R_{2N+2M})=(U_1,\ldots,U_N;V_1,\ldots,V_M;W_1,\ldots,W_{N+M})$.
\\
\indent Some linearly degenerate
non-homogeneous Riemann invariants forms, associated with commuting
quadratic Hamiltonians and the Killing vector fields  of the given
metric, were investigated in \cite{Fo1, Fo2}. However, in our case equation (\ref{N-Rie-inv})
is obviously not linearly degenerate.

 \textbf{Remark.} For the type of non-homogeneous Riemann
invariant form
\begin{equation}
 \pa_{t_n}R^i=\La_n^i(\mathbf{R})R^i_x + Q_n^i(\mathbf{R}),
\label{NRI}
\end{equation}
the requirements of the commutativity are equivalent to the following restrictions
on their characteristic speeds and non-homogeneous terms
(see appendix A)
\[
\frac{\pa_j\La_n^i}{\La_n^j-\La_n^i} =\frac{\pa_j\La_m^i}{\La_m^j-\La_m^i}, \quad
\frac{\pa_jQ_n^i}{Q_n^j} =\frac{\pa_jQ_m^i}{Q_m^j}, \quad
\frac{Q_n^j}{\La_n^j-\La_n^i} =\frac{Q_m^j}{\La_m^j-\La_m^i},
\quad i\neq j, \quad n\neq m.
\]
where $\pa_i\equiv \pa/\pa R^i$.
\\
%%%%%%%%%%%%%%%%%%%%%%%%%%%%%%%%%%%
%
\textbf{Example 1.}
%
%%%%%%%%%%%%%%%%%%%%%%%%%%%%%%%%%%%
$(N,M)=(1,1)$ reduction. In this case,
\begin{eqnarray*}
 \mL &=& p + \log(1-U/p), \\
 \mM &=& \sum_{n=1}^{\infty}nt_n\mL^{n-1} + \log(1-V/p).
\end{eqnarray*}
Comparing to the expansion of (\ref{L},\ref{M}) we have $u_n=-U^n/n$ for $n\geq 1$
and $v_1=-V, v_2=-V^2/2, v_3=UV-V^3/3$, etc.
These transformations allow us to get $A_n, B_n$ (by Eqs.(\ref{A}), (\ref{B}))
which correspond to the reduced system.
The $G$ function is given by
\begin{equation}
 G = \frac{p(p-U)(p-V)}{\prod_{i=1}^3(p-W_i)},
\label{G11}
\end{equation}
where $W_i$ satisfy
\begin{equation}
\sum_{i=1}^3W_i=U+V+V_x,\quad
\sum_{\scriptstyle i,j=1 \atop\scriptstyle (i>j)}^3 W_iW_j=U+UV+UV_x, \quad
\prod_{i=1}^3W_i= UV+UV_x-U_xV.
\label{Weq}
\end{equation}
Notice that (\ref{G11}) is not coincident with that in (\ref{G-Wb}),
there is one more root of $p=0$ to be considered.
By (\ref{Geq}), it turns out that the evaluation of $p=0$ gives an additional condition, namely
\begin{equation}
 UVB_n(p=0)=0, \quad \forall n\geq 1.
\end{equation}
There are two simple cases:
(i) $V=0,U\neq 0$, (ii) $V\neq 0,U=0$.
One can easily deduce considering $t_2$-flow of (\ref{N-Rie-inv})
that case (i) is a trivial reduction.
For the case (ii), we have the fact that $B_n(U=0)=0$ for $n\geq 1$,
and Eq.(\ref{Weq}) will reveal us the only one relation: $V_x=W-V$.
To this end, system (\ref{N-Rie-inv}) reduces to the type of homogeneous one in (\ref{NRI})
with $Q_n^i=0$, namely
\begin{equation}
 \pa_{t_n}R^i = \Lambda_n^i(\mathbf{R})\pa_xR^i,
\label{VWeq}
\end{equation}
where $\mathbf{R}=(R^1,R^2)=(V,W)$ and the characteristic speeds $\Lambda_n^i=A_n(p=R^i,U=0)$.
For instance, for $t_2=y$ flow, we have $A_2(U=0)=2p+2V_x=2p+2(W-V)$,
then Eq.(\ref{VWeq}) becomes
\begin{equation}
\left(\ba{c} V\\W \ea\right)_y
= \left(\ba{cc} 2W & 0\\0&4W-2V \ea\right)\left(\ba{c}  V\\W \ea\right)_x.
\label{(1,1)-t2}
\end{equation}
For $t_3=t$ flow, we derive
$A_3(U=0)=3p^2+3(W-V)p+3(W-V)^2+3V(W-V)$, thus
\begin{equation}
\left(\ba{c} V\\W \ea\right)_t
= \left(\ba{cc} 3W^2 & 0\\0 & 9W^2-6VW \ea\right)\left(\ba{c} V\\W \ea\right)_x.
\label{(1,1)-t3}
\end{equation}
From the two nontrivial flows (\ref{(1,1)-t2}), (\ref{(1,1)-t3}),
we readily obtain the following set of hodograph equation
\begin{eqnarray}
 x+2Wy+3W^2t &=& \hat{F}(V,W), \no\\
 x+(4W-2V)y+(9W^2-6VW)t &=& \hat{G}(V,W),
\label{hodo-1}
\end{eqnarray}
where $\hat{F}$ and $\hat{G}$ satisfy the linear equations
\begin{eqnarray*}
(W-V)\hat{G}_V &=& \hat{F}-\hat{G}, \\
(W-V)\hat{F}_W &=& \hat{G}-\hat{F}.
\end{eqnarray*}
Dividing these two equations for $V\neq W$ we get $\hat{G}_V=-\hat{F}_W$.
It follows that there exists a function $\phi$ such that $\hat{F}=\phi_V,\hat{G}=-\phi_W$,
whence $\phi$ satisfies the defining equation
\begin{equation}
 (V-W)\phi_{VW} = \phi_V+\phi_W.
\label{phi-eq}
\end{equation}
Eq. (\ref{phi-eq}) has a general solution of the form
\[
 \phi = (V-W)\left(f(W)+\int\frac{g(V)}{(V-W)^2}\,dV\right),
\]
where $f(W)$ and $g(V)$ are arbitrary functions of $W$ and $V$, respectively.
Choosing, for example $f(W)=W^3$, $g(V)=\mbox{Const.}$, then we have
$\hat{F}=W^3$ and $\hat{G}=-3VW^2+4W^3$.
Substituting back into the hodograph equation (\ref{hodo-1}) we solve
\begin{eqnarray*}
V=W &=& \frac{1}{6}h^{-1/3}(24y+36t^2+6th^{1/3}+h^{2/3}), \\
  h &=& 216yt+108x+216t^3+12\sqrt{-96y^3-108y^2t^2+324ytx+81x^2+324xt^3},
\end{eqnarray*}
which satisfies the $t_2$- and $t_3$-flows (\ref{(1,1)-t2}),
(\ref{(1,1)-t3}). However, $V=W$ contradicts to the relation
$W=V+V_x$ and $V$ does not satisfy equation
(\ref{Pavlov}). Actually, from equation (\ref{phi-eq}) we
can see that when $V=W$ one can get  $\hat{F}=\hat {G}$. Then we
obtain all the solutions will satisfy $V=W$. Consequently, there
is no (1,1)- reduction. Similar considerations can show that there
are no (1,2)- and (2,1)- reductions, either.
\\
%%%%%%%%%%%%%%%%%%%%%%%%%%%%%%%%%%%
%
\textbf{Example 2.}
%
%%%%%%%%%%%%%%%%%%%%%%%%%%%%%%%%%%%
$(N,M)=(2,2)$ reduction. In this case,
\begin{eqnarray*}
 \mL &=& p+\ep_1\log\frac{p-U_1}{p-U_2}, \\
 \mM &=& \sum_{n=1}^{\infty}nt_n\mL^{n-1} + \delta_1\log\frac{p-V_1}{p-V_2}.
\end{eqnarray*}
For simplicity, we set $\ep_1=\delta_1=1$.
Comparing to the expansion of (\ref{L},\ref{M}), we have $u_n=(U_2^n-U_1^n)/n$ for $n\geq 1$
and $v_1=V_2-V_1, v_2=(V_2^2-V_1^2)/2, v_3=u_1v_1+(V_2^3-V_1^3)/3,\ldots$.
Now we expand the hierarchy flow of $U_i$, $V_i$ and $W_i$ up to $t_2=y, t_3=t$.
From (\ref{G-Wb}) with $(N,M)=(2,2)$ we have
\[
G=\frac{\prod_{i=1}^{2}(p-U_i)\prod_{j=1}^{2}(p-V_j)}{\prod_{k=1}^{4}(p-W_k)}.
\]
where $W_i$ satisfy
\begin{eqnarray}
\sum_{i=1}^4W_i
&=& U_1+U_2+V_1+V_2+V_{1x}-V_{2x},
\label{Weq1}\\
\sum_{\scriptstyle i,j=1 \atop\scriptstyle (i>j)}^4 W_iW_j
&=& U_1-U_2+U_1U_2+V_1V_2+V_{1x}V_2-V_1V_{2x} \no\\[-0.6cm]
 && +(U_1+U_2)(V_1+V_2+V_{1x}-V_{2x}),
\label{Weq2} \\
\sum_{i=1}^4W_i^{-1}\prod_{j=1}^4W_j
&=& (U_1+U_2)(V_1V_2+V_{1x}V_2-V_1V_{2x})+(U_{2x}-U_{1x})(V_1-V_2) \no\\[-0.3cm]
 &&  +(U_1-U_2+U_1U_2)(V_1+V_2+V_{1x}-V_{2x}),
\label{Weq3}\\
\prod_{i=1}^4W_i
&=& (U_1-U_2+U_1U_2)(V_1V_2+V_{1x}V_2-V_1V_{2x}) \no\\[-0.3cm]
 &&  -(V_1-V_2)(U_{1x}U_2-U_1U_{2x}),
\label{Weq4}
\end{eqnarray}
from which, one can substitute into $A_n,B_n$ to eliminate $U_{ix}, V_{ix}$, etc.
For $n=2$, using (\ref{Weq1}), (\ref{Weq3}) we have
\begin{eqnarray*}
A_2(p)
 &=& 2p+2(V_{1x}-V_{2x})
  = 2p + 2\Big(-U_1-U_2-V_1-V_2+\sum_{i=1}^4W_i\Big), \\
 &=& 2(p-R_1-R_2-R_3-R_4+R_5+R_6+R_7+R_8),
\end{eqnarray*}
and the non-homogeneous term
\begin{eqnarray*}
&& B_2(p)
  =  2(U_{2x}-U_{1x}), \\
 &&= \frac{2}{V_1-V_2}\Bigg[\sum_{i=1}^4W_i^{-1}\prod_{j=1}^4W_j
      + (U_1+U_2)\bigg(U_1-U_2+U_1U_2-\sum_{i>j}^4W_iW_j\bigg)  \\
  && \quad + \Bigg.\bigg(U_1+U_2-\sum_{i=1}^4W_i\bigg)\bigg(U_1-U_2+U_1U_2-(U_1+U_2)^2\bigg) \Bigg],\\
 &&= \frac{2}{R_3-R_4}\Big[(R_1+R_2-R_5-R_6-R_7-R_8)(R_1-R_2-R_1R_2-R_1^2-R_2^2) \Big. \\
  && \quad +(R_1+R_2)(R_1-R_2+R_1R_2-R_5R_6-R_5R_7-R_5R_8-R_6R_7-R_6R_8-R_7R_8) \\
  && \quad +R_5R_6R_7+R_6R_7R_8+R_7R_8R_5+R_8R_5R_6\Big].
\end{eqnarray*}
Then the $t_2=y$ flow in (\ref{N-Rie-inv}) is now read
\begin{equation}
 \pa_yR_i = 2(R_i-R_1-R_2-R_3-R_4+R_5+R_6+R_7+R_8)R_{ix} + B_2.
\label{Ri-y}
\end{equation}
For $n=3$, Eq.(\ref{N-Rie-inv}) becomes
\begin{eqnarray*}
 \frac{\pa R_i}{\pa t}
&=& A_3(p=R_i)R_{ix} + B_3(p=R_i), \\
&=& \left[3p^2+3(V_{1x}-V_{2x})p+6(U_2-U_1)
          +3(V_{1x}-V_{2x})^2+\frac{3}{2}(V_1^2-V_2^2)_x\right]_{p=R_i}R_{ix}\\
 && +\left[3(U_{2x}-U_{1x})p-3(U_{2x}-U_{1x})(V_{2x}-V_{1x})+\frac{3}{2}(U_2^2-U_1^2)_x\right]_{p=R_i}\\
&=& \left[3R_i^2+3R_i(V_{1x}-V_{2x})
    +6(U_2-U_1)+3(V_{1x}-V_{2x})^2+\frac{3}{2}(V_1^2-V_2^2)_x\right]R_{ix} \\
 && +3R_i(U_{2x}-U_{1x})-3(U_{2x}-U_{1x})(V_{2x}-V_{1x})+3(U_2U_{2x}-U_1U_{1x}).
\end{eqnarray*}
Using Eqs.(\ref{Weq1})--(\ref{Weq4}) we arrive
\begin{eqnarray*}
\frac{\pa R_i}{\pa t}
&=&  3R_{ix}\Bigg[U_2-U_1+R_i\Big(R_i-U_1-U_2-V_1-V_2+\sum_{i=1}^4W_i\Big) \\
&&   +U_1U_2-V_1V_2-V_1^2-V_2^2 -\sum_{i>j}^4W_iW_j
     +\Big(U_1+U_2+V_1+V_2-\sum_{i=1}^4W_i\Big)^2  \\
&&   \qquad+(U_1+U_2+V_1+V_2)\Big(-U_1-U_2+\sum_{i=1}^4W_i\Big)\Bigg] \\
&&   +\frac{3R_i}{V_1-V_2}\Bigg[(U_1+U_2)\Big(U_1-U_2+U_1U_2-\sum_{i>j}^4W_iW_j\Big)
     +\sum_{i=1}^4W_i^{-1}\prod_{j=1}^4W_j  \\
&&   \quad + \Big(U_1+U_2-\sum_{i=1}^4W_i\Big)
             \Big(U_1-U_2+U_1U_2-(U_1+U_2)^2\Big) \Bigg]+\\
 &&  +3\Big(-U_1-U_2-V_1-V_2+\sum_{i=1}^4W_i\Big)\times \\
 &&  \times\frac{1}{V_1-V_2}\Bigg(\Big(U_1+U_2-\sum_{i=1}^4W_i\Big)\Big(U_1-U_2+U_1U_2-(U_1+U_2)^2\Big) \\
 && \qquad+(U_1+U_2)\Big(U_1-U_2+U_1U_2-\sum_{i>j}W_iW_j\Big)+\sum_{i-1}^4W_i^{-1}\prod_{j=1}^4W_j\Bigg) \\
&&  +\frac{3}{V_1-V_2}\Bigg((U_1+U_2)\sum_{i-1}^4W_i^{-1}\prod_{j=1}^4W_j
    +(U_1+U_2)^2\Big(U_1-U_2+U_1U_2-\sum_{i>j}W_iW_j\Big) \\
&&  -\prod_{i=1}^4W_i
    +(U_1+U_2)\Big(U_1+U_2-\sum_{i=1}^4W_i\Big)\Big(U_1-U_2+U_1U_2-(U_1+U_2)^2\Big) \\
&&  +(U_1-U_2+U_1U_2)\Big(\sum_{i>j}^4W_iW_j-(U_1-U_2)-U_1U_2+(U_1+U_2)^2
    -(U_1+U_2)\sum_{i=1}^4W_i\Big)\Bigg).
\end{eqnarray*}
Expressing in terms of $R_i,i=1,\ldots,8$, we get
\begin{eqnarray*}
\frac{\pa R_i}{\pa t}
&=& 3R_{ix}\Bigg[R_2-R_1+R_i(R_i-R_1-R_2-R_3-R_4+R_5+R_6+R_7+R_8) \\
&&\quad  +R_1 R_2+R_3 R_4+R_5 R_6+R_5 R_7+R_5 R_8+R_6 R_7+R_6 R_8+R_7 R_8\\
&&\quad  +R_1 R_3+R_1 R_4-R_1 R_5-R_1 R_6-R_1 R_7-R_1 R_8 +R_2 R_3+R_2 R_4 \\
&&\quad  -R_2 R_5-R_2 R_6-R_2 R_7-R_2 R_8-R_3 R_5-R_3 R_6-R_3 R_7-R_3 R_8 \\
&&\quad  -R_4 R_5-R_4 R_6-R_4 R_7-R_4 R_8 +R_5^2+R_6^2+R_7^2+R_8^2\Bigg] \\
&& +\frac{3R_i}{R_3-R_4}\Big(2 R_1^2-2 R_2^2- R_1R_5-R_1R_6 -R_1R_7 -R_1R_8+R_2R_5 +R_2R_6  \\
&&\quad + R_2R_7+ R_2R_8-R_1^3-R_2^3-R_1^2 R_2 +R_1^2R_5+R_1^2R_6+R_1^2R_7+R_1^2R_8  \\
&&\quad -R_1 R_2^2+ R_2^2R_5 +R_2^2R_6 +R_2^2R_7 +R_2^2R_8 + R_1 R_2R_5+ R_1 R_2R_6\\
&&\quad + R_1 R_2R_7+ R_1 R_2R_8-R_1 R_5 R_6-R_1 R_5 R_7-R_1 R_5 R_8-R_1 R_6 R_7\\
&&\quad -R_1 R_6 R_8-R_1 R_7 R_8-R_2 R_5 R_6-R_2 R_5 R_7-R_2 R_5 R_8-R_2 R_6 R_7\\
&&\quad -R_2 R_6 R_8-R_2 R_7 R_8+R_5 R_6 R_7+R_6 R_7 R_8+R_7 R_8 R_5+R_8 R_5 R_6\Big)+ \\
&& +\frac{3}{R_3-R_4}\Big(-R_1^2-R_2^2+2 R_1 R_2+R_1^3-R_2^3-R_1^2 R_2+R_1 R_2^2-2 R_1^2R_3 +2 R_2^2R_3\\
&&\quad  -2 R_1^2R_4 +2 R_2^2R_4+ R_1^2R_5- R_2^2R_5+ R_1^2R_6- R_2^2R_6+ R_1^2R_7- R_2^2R_7  \\
&&\quad  + R_1^2R_8- R_2^2R_8-R_1R_5^2 +R_2R_5^2+R_1R_6^2 -R_2R_6^2 +R_1R_7^2 - R_2R_7^2 \\
&&\quad  + R_1R_8^2- R_2R_8^2+R_1^3 R_2+R_1^3R_3 +R_1^3R_4 -R_1^3R_5 -R_1^3R_6 -R_1^3R_7 -R_1^3R_8   \\
&&\quad  +R_1 R_2^3+ R_2^3R_3+ R_2^3R_4 -R_2^3R_5- R_2^3R_6- R_2^3R_7-R_2^3R_8 \\
&&\quad  +R_1^2 R_2^2+ R_1^2R_5^2+ R_2^2R_5^2- R_1^2R_6^2
         - R_2^2R_6^2- R_1^2R_7^2- R_2^2R_7^2- R_1^2R_8^2- R_2^2R_8^2\\
&&\quad  +R_1 R_5 R_6+R_1 R_5 R_7+R_1 R_5 R_8+3 R_1 R_6 R_7+3 R_1 R_6 R_8+3 R_1 R_7 R_8\\
&&\quad  -R_2 R_5 R_6-R_2 R_5 R_7-R_2 R_5 R_8-3 R_2 R_6 R_7-3 R_2 R_6 R_8-3 R_2 R_7 R_8\\
&&\quad  +R_3 R_5 R_1-R_3 R_5 R_2-R_3 R_6 R_1+R_3 R_6 R_2-R_3 R_7 R_1+R_3 R_7 R_2 \\
&&\quad  -R_3 R_8 R_1+R_3 R_8 R_2+R_4 R_5 R_1-R_4 R_5 R_2-R_4 R_6 R_1+R_4 R_6 R_2 \\
&&\quad  -R_4 R_7 R_1+R_4 R_7 R_2-R_4 R_8 R_1+R_4 R_8 R_2-2 R_5 R_1^2 R_2-2 R_6 R_1^2 R_2 \\
&&\quad  -2 R_7 R_1^2 R_2-2 R_1 R_6 R_2^2-2 R_1 R_7 R_2^2-2 R_1 R_8 R_2^2-2 R_8 R_1^2 R_2-2 R_1^2 R_6 R_7\\
&&\quad  -2 R_1^2 R_6 R_8-2 R_1^2 R_7 R_8-2 R_2^2 R_6 R_7-2 R_2^2 R_6 R_8 -2 R_2^2 R_7 R_8-2 R_1 R_5 R_2^2\\
&&\quad  +R_5^2 R_6 R_7+R_7 R_8 R_5^2+R_8 R_5^2 R_6+R_5^2 R_1 R_2-R_1 R_5^2 R_6-R_1 R_5^2 R_7-R_1 R_5^2 R_8\\
&&\quad  -R_2 R_5^2 R_6-R_2 R_5^2 R_7-R_2 R_5^2 R_8+R_5 R_6^2 R_7+R_6^2 R_7 R_8+R_8 R_5 R_6^2-R_6^2 R_1 R_2\\
&&\quad  -R_1 R_5 R_6^2-R_1 R_6^2 R_7-R_1 R_6^2 R_8-R_2 R_5 R_6^2-R_2 R_6^2 R_7-R_2 R_6^2 R_8+R_5 R_6 R_7^2\\
&&\quad  +R_6 R_7^2 R_8+R_7^2 R_8 R_5-R_7^2 R_1 R_2-R_1 R_5 R_7^2-R_1 R_6 R_7^2-R_1 R_7^2 R_8-R_2 R_5 R_7^2\\
&&\quad  -R_2 R_6 R_7^2-R_2 R_7^2 R_8+R_6 R_7 R_8^2+R_7 R_8^2 R_5+R_8^2 R_5 R_6-R_8^2 R_1 R_2-R_1 R_5 R_8^2\\
&&\quad  -R_1 R_6 R_8^2-R_1 R_7 R_8^2-R_2 R_5 R_8^2-R_2 R_6 R_8^2-R_2 R_7 R_8^2+R_3 R_1^2 R_2+R_3 R_1 R_2^2\\
&&\quad  -R_3 R_5 R_1^2-R_3 R_5 R_2^2+R_3 R_6 R_1^2+R_3 R_6 R_2^2+R_3 R_7 R_1^2+R_3 R_7 R_2^2+R_3 R_8 R_1^2\\
&&\quad  +R_3 R_8 R_2^2+R_4 R_1^2 R_2+R_4 R_1 R_2^2-R_4 R_5 R_1^2-R_4 R_5 R_2^2+R_4 R_6 R_1^2+R_4 R_6 R_2^2\\
&&\quad  +R_4 R_7 R_1^2+R_4 R_7 R_2^2+R_4 R_8 R_1^2+R_4 R_8 R_2^2-3 R_1 R_5 R_6 R_7-3 R_1 R_6 R_7 R_8 \\
&&\quad  -3 R_1 R_5 R_7 R_8-3 R_1 R_5 R_6R_8-3 R_2 R_5 R_6 R_7-3 R_2 R_6 R_7 R_8+3 R_5 R_6 R_7 R_8 \\
&&\quad  -3 R_2 R_7 R_8 R_5-3 R_2 R_8 R_5 R_6+R_1 R_2 R_5 R_6+R_1 R_2 R_5 R_7+R_1 R_2 R_5 R_8\\
&&\quad  -R_1 R_2 R_6 R_7-R_1 R_2 R_6 R_8-R_1 R_2 R_7 R_8-R_3 R_5 R_6 R_7-R_3 R_6 R_7 R_8\\
&&\quad  -R_3 R_7 R_8 R_5-R_3 R_8 R_5 R_6-R_3 R_5 R_1 R_2+R_3 R_6 R_1 R_2+R_3 R_7 R_1 R_2\\
&&\quad  +R_3 R_8 R_1 R_2+R_3 R_1 R_5 R_6+R_3 R_1 R_5 R_7+R_3 R_1 R_5 R_8+R_3 R_1 R_6 R_7\\
&&\quad  +R_3 R_1 R_6 R_8+R_3 R_1 R_7 R_8+R_3 R_2 R_5 R_6+R_3 R_2 R_5 R_7+R_3 R_2 R_5 R_8\\
&&\quad  +R_3 R_2 R_6 R_7+R_3 R_2 R_6 R_8+R_3 R_2 R_7 R_8-R_4 R_5 R_6 R_7-R_4 R_6 R_7 R_8\\
&&\quad  -R_4 R_7 R_8 R_5-R_4 R_8 R_5 R_6-R_4 R_5 R_1 R_2+R_4 R_6 R_1 R_2+R_4 R_7 R_1 R_2\\
&&\quad  +R_4 R_8 R_1 R_2+R_4 R_1 R_5 R_6+R_4 R_1 R_5 R_7+R_4 R_1 R_5 R_8+R_4 R_1 R_6 R_7\\
&&\quad  +R_4 R_1 R_6 R_8+R_4 R_1 R_7 R_8+R_4 R_2 R_5 R_6+R_4 R_2 R_5 R_7+R_4 R_2 R_5 R_8\\
&&\quad  +R_4 R_2 R_6 R_7+R_4 R_2 R_6 R_8+R_4 R_2 R_7 R_8
\Big).
\end{eqnarray*}
\section{Concluding Remarks}
In this article, we investigate the Manakov-Santini equation starting from
Lax-Sato formulation of associated hierarchy and obtain equations (\ref{J0eq}),
(\ref{Geq}), which generalize the results of \cite{MS02}. From
these, one can introduce new coordinates (\ref{G-Wb} ) such that
the non-hydrodynamic evolution (\ref{y-flow}), (\ref{t-flow}) of
waterbag reduction transforms to non-homogeneous Riemann invariants form
of hydrodynamic type (\ref{N-Rie-inv}). The equation
(\ref{N-Rie-inv}) is not linearly degenerate. Hence  the
generalization of \cite{Fo1, Fo2} from linearly degenerate  case
to the general one could be interesting. Also, the solution
structures of (\ref{N-Rie-inv}) having infinite symmetries should
be investigated. These issues will be published elsewhere.

%%%%%%%%%%%%%%%%%%%%%%%%%%%%%%%%%%%%%%%%%%%%%%%%%%%%%%%%
\renewcommand{\theequation}{A.\arabic{equation}}
\section*{Appendix}
\appendix
\setcounter{equation}0
\section{Commutability properties of the non-homogeneous diagonal system}
%%%%%%%%%%%%%%%%%%%%%%%%%%%%%%%%%%%%%%%%%%%%%%%%%%%%%%%%
We start from the commutability of (\ref{NRI}) by $\pa_m\pa_nR^i=\pa_n\pa_mR^i$:
\begin{eqnarray*}
&& \pa_m\pa_nR^i \\
&&= \pa_m(\La_n^iR^i_x)+\pa_mQ_n^i, \\
&&= \sum_j(\pa_j\La_n^i)(\pa_mR^j)R^i_x + \La_n^i\pa_x(\pa_mR^i)
    + \sum_j(\pa_jQ_n^i)(\pa_mR^j), \\
&&= \sum_j(\pa_j\La_n^i)(\La_m^jR^j_x+Q_m^j)R^i_x
     + \La_n^i\pa_x(\La_m^iR^i_x+Q_m^i)
    + \sum_j(\pa_jQ_n^i)(\La_m^jR^j_x+Q_m^j), \\
&&= \sum_j(\pa_j\La_n^i)(\La_m^jR^j_x+Q_m^j)R^i_x
    + \La_n^i\Big(\sum_j(\pa_j\La_m^i)R^j_xR^i_x + \La_m^iR^i_{xx}
    + \sum_j(\pa_jQ_m^i)R^j_x\Big) \\
&&\quad  + \sum_j(\pa_jQ_n^i)(\La_m^jR^j_x+Q_m^j), \\
&&= \sum_j\Big[(\pa_j\La_n^i)\La_m^j+(\pa_j\La_m^i)\La_n^i\Big]R^j_xR^i_x
    +\sum_j(\pa_j\La_n^i)Q_m^jR^i_x + \La_n^i\La_m^iR^i_{xx} \\
&&\quad  + \sum_j\Big[(\pa_jQ_m^i)\La_n^i+(\pa_jQ_n^i)\La_m^j\Big]R^j_x
      + \sum_j(\pa_jQ_n^i)Q_m^j.
\end{eqnarray*}
Similarly,
\begin{eqnarray*}
\pa_n\pa_mR^i
&=& \pa_n(\La_m^iR^i_x)+\pa_nQ_m^i, \\
&=& \sum_j\Big[(\pa_j\La_m^i)\La_n^j+(\pa_j\La_n^i)\La_m^i\Big]R^j_xR^i_x
    +\sum_j(\pa_j\La_m^i)Q_n^jR^i_x + \La_m^i\La_n^iR^i_{xx} \\
&&\quad  + \sum_j\Big[(\pa_jQ_n^i)\La_m^i+(\pa_jQ_m^i)\La_n^j\Big]R^j_x
      + \sum_j(\pa_jQ_m^i)Q_n^j.
\end{eqnarray*}
Then, $\pa_m\pa_nR^i=\pa_n\pa_mR^i$ provide the following compatibility conditions:
\begin{itemize}
\item [(i)] Taking the coefficients of $R^j_xR^i_x$ we have
\[
 (\pa_j\La_n^i)\La_m^j+(\pa_j\La_m^i)\La_n^i = (\pa_j\La_m^i)\La_n^j+(\pa_j\La_n^i)\La_m^i,
\]
which implies
\begin{equation}
 \frac{\pa_j\La_n^i}{\La_n^j-\La_n^i} = \frac{\pa_j\La_m^i}{\La_m^j-\La_m^i}.
\label{c1}
\end{equation}
\item [(ii)] Taking the coefficients of $R^i_x$ we have
\[
 (\pa_j\La_n^i)Q_m^j = (\pa_j\La_m^i)Q_n^j.
\]
Combining (\ref{c1}), the above equation can be written as
\begin{equation}
 \frac{Q_n^j}{\La_n^j-\La_n^i} = \frac{Q_m^j}{\La_m^j-\La_m^i}.
\label{c2}
\end{equation}
\item [(iii)] Taking the coefficients of $R^j_x$ we get
\begin{equation}
 \frac{\pa_jQ_n^i}{\La_n^j-\La_n^i} = \frac{\pa_jQ_m^i}{\La_m^j-\La_m^i}.
\label{c3}
\end{equation}
\item [(iv)] The zero-th term of $\pa_m\pa_nR^i=\pa_n\pa_mR^i$ give us
\begin{equation}
 \frac{\pa_jQ_n^i}{Q_n^j} = \frac{\pa_jQ_m^i}{Q_m^j}.
\label{c4}
\end{equation}
\end{itemize}
Notice that according to (\ref{c2}), equation (\ref{c4}) is equivalent to (\ref{c3}).
To summarize, we have three compatibility conditions (\ref{c1}), (\ref{c2}) and (\ref{c4}).

\section*{Acknowledgments} The first author (LVB) is grateful to National Defense University
(Taoyuan, Taiwan), where a part of this work was done, for
hospitality. This research was particularly supported by the
Russian-Taiwanese grant 95WFE0300007 (RFBR grant 06-01-89507) and
NSC-95-2923-M-606-001-MY3. LVB was also supported in part by RFBR
grant 07-01-00446.

%%%%%%%%%%%%%%%%%%%%%%%%%%%%%%%%%%%%%%%%%%%%%%%%%%%%%%%%%

\end{document}